\documentclass[12pt,a4paper]{article}
\usepackage{fullpage}

\usepackage{amsmath}
\usepackage{mathtools}
\usepackage{amsthm}
\usepackage{amssymb}
\usepackage{amsfonts}
\usepackage{mleftright}
\usepackage{enumitem}  
\usepackage{graphicx}
\usepackage{systeme}
\usepackage{ltablex,booktabs,ragged2e,caption}

\usepackage[table]{xcolor}
\theoremstyle{definition}
\newtheorem{example}{Example}

\makeatletter
\newcommand{\longdash}[1][2em]{%
	\makebox[#1]{$\m@th\smash-\mkern-7mu\cleaders\hbox{$\mkern-2mu\smash-\mkern-2mu$}\hfill\mkern-7mu\smash-$}}

\makeatletter
\renewcommand*\env@matrix[1][\arraystretch]{%
	\edef\arraystretch{#1}%
	\hskip -\arraycolsep
	\let\@ifnextchar\new@ifnextchar
	\array{*\c@MaxMatrixCols c}}
\makeatother

\begin{document}

\title{Private Information Retrieval using Product-Matrix Minimum Storage Regenerating Codes}

\author{Chatdanai Dorkson\footnote{Department of Mathematics, Royal Holloway, University of London, Egham, Surrey, TW20 0EX, United Kingdom. Chatdanai.Dorkson.2016@rhul.ac.uk}, Siaw-Lynn Ng\footnote{Information Security Group, Royal Holloway, University of London, Egham, Surrey, TW20 0EX, United Kingdom. S.Ng@rhul.ac.uk}}
\date{\today}
\maketitle

\begin{abstract}
	Private Information Retrieval (PIR) schemes allow a user to retrieve a record from the server without revealing any information on which record is being downloaded. In this paper, we consider PIR schemes where the database is stored using Minimum Storage Regenerating (MSR) codes which is a class of optimal regenerating codes providing efficient repair when a node failure occurs in the system. We analyse the relationship between the costs of privacy, storage and repair, and also construct an explicit PIR scheme that uses the MSR codes from \cite{matrix} to achieve the optimal curve of the trade-off.
\end{abstract}

%\keywords{Private Information Retrieval; MSR Codes; Regenerating Codes}

\vspace{13pt}
\section{Introduction}
\label{sec1}

Private information retrieval (PIR) schemes allow a user to download a record from the database without revealing any information on which record the user wants to retrieve. In the classical setting for PIR \cite{Original}, the entire database is replicated among $n$ non-communicating nodes. Since the replication-based PIR results in high storage cost, this motivates the use of erasure codes which means that each node stores only a fraction of the entire database, hence reducing the storage cost while achieving reliability. We call this \textit{code-based PIR schemes}. 

Shah et al. \cite{extra} are the first to explore the work on coded databases. They give an explicit PIR scheme where only one extra bit of download is sufficient to retrieve the desired record. However, this construction requires the number of nodes in the system to be exponential in the number of records. As the result of the exponential storage cost, they provide another PIR scheme using the product-matrix Minimum Bandwidth Regenerating (MBR) codes \cite{matrix}. The storage overhead in this scheme is linear and the total download is just a small factor away from optimality. 

In \cite{tho}, Chan et al. propose retrieval schemes for a general class of linear storage codes, and the trade-off between storage cost and retrieval cost in the context of their proposed PIR schemes is analysed. Subsequently in \cite{MDS}, Tajeddine and Rouayheb give an explicit PIR scheme using MDS codes which achieves the optimal retrieval cost for linear schemes (\cite{tho}). Kumar et al. \cite{arbitrary} show that an arbitrary systematic linear storage code of rate $ > 1/2$ can be used in the PIR scheme, and, interestingly, numerical results show that the optimal PIR protocal can be achieved by using locally repairable codes (LRCs) \cite{LRCs} and Pyramid codes \cite{pyramid} which have more efficient repair property. In \cite{capacity}, Banawan and Ulukus consider the problem of capacity - the maximum of the retrieval rates over all possible PIR schemes. They analyse the capacity of the PIR scheme using MDS codes as in \cite{MDS} which improves the results in \cite{tho} and \cite{MDS} and gives the best-known upper bound on the retrieval rate.

In this paper, we consider the PIR problem in coded databases. Specifically, we consider the situation where the database is stored using minimum storage regenerating (MSR) codes (\cite{matrix}). The interesting property of regenerating codes is that such codes optimally trade the bandwidth needed to repair a failed node with the storage overhead, hence our scheme obtains a lower repair cost compared to schemes that use MDS codes. 

The organisation of this paper is as follows. In Section 2, we briefly recall the product-matrix Minimum Storage Regenerating codes from \cite{matrix}. Section 3 provides the system model of PIR scheme using product-matrix MSR codes. The decodability condition and trade-off analysis between the storage cost and the retrieval cost in the system are given in Section 4. Section 5 then presents a motivating example and the general construction of our PIR scheme. Lastly, we discuss the efficiency of our scheme in Section 6.

\vspace{15pt}
\section{Product-Matrix {MSR} Codes}

\subsection{Minimum Storage Regenerating ({MSR}) Codes}

An $(n,k,r,\alpha,\beta,B)$ \textit{regenerating code} \cite{dimakis} is defined to be a distributed storage code storing the database of size $B$ among $n$ nodes where each node stores $\alpha$ symbols satisfying two properties: (i)(recovery) The entire database can be recovered from the data stored in any $k$ nodes; (ii)(repair) If one of the storage nodes fails, then a newcomer node connects to some set of $r$ remaining nodes where $k<r<n$, and downloads $\beta$ symbols from each of these $r$ nodes in order to regenerate $\alpha$ symbols in such a way that we can perform (i) and (ii) again when another node failure occurs. 

The total amount of $r\beta$ symbols downloaded for regenerating is called the \textit{repair bandwidth}, and typically the repair bandwidth is smaller than the size of the whole database. There are various repair models, but for PIR we focus on the exact repair model, where a newcomer node will regenerate the same data as was stored in the failed node in order to maintain the initial state of the storage nodes. 

In \cite{YWu}, the parameters of a regenerating code is shown to necessarily satisfy $$ B \leq \sum_{i=0}^{k-1} \min\{\alpha,(r-i)\beta\},$$ 
and the achievable trade-off between storage overhead and repair bandwidth is characterised by fixing the repair bandwidth, and then deriving the minimum $\alpha$ which satisfies the above equation. An interesting extremal point on the optimal trade-off curve is the \textit{minimum storage regeneration} (MSR) point which minimises storage overhead first and then minimises repair bandwidth. It can be shown that the MSR point is achieved by $$(\alpha_{MSR},\beta_{MSR}) = \bigg(\frac{B}{k}, \frac{B}{k(r-k+1)}\bigg),$$ and \textit{MSR codes} are $(n,k,r,\alpha,\beta,B)$ regenerating codes that satisfies the above equation. 

\subsection{The Product-Matrix {MSR} Codes (\cite{matrix})}

Under the product-matrix framework, each codeword is represented by an $(n \times \alpha)$ \textit{code matrix} $C$ which is the product $$C = \Psi \cdot M$$ of an $(n \times r)$ \textit{encoding matrix} $\Psi$ and an $(r \times \alpha)$ \textit{message matrix} $M$. The message matrix $M$ contains the $B$ message symbols. In the code matrix $C$, row $i$ consists of the $\alpha$ encoded symbols stored by node $i$ for each $i \in [n]$.

In \cite{matrix}, Rashmi, Shah and Kumar give an explicit construction for the MSR code with $r=2k-2$, so the parameters are $(n,k,r,\alpha,\beta,B) = (n,k,2k-2,k-1,1,k(k-1))$ where $n>2k-2$ using the product-matrix framework. First, they let the encoding matrix $\Psi$ be any $(n \times r)$ matrix given by
$$ \Psi = \begin{bmatrix} 
\Phi & \Lambda\Phi \\
\end{bmatrix}$$
where $\Phi$ is an $(n \times \alpha)$ matrix and $\Lambda$ is an $(n \times n)$ diagonal matrix such that (i) any $r$ rows of $\Psi$ are linearly independent, (ii) any $\alpha$ rows of $\Phi$ are linearly independent, (iii) the $n$ diagonal elements of $\Lambda$ are all distinct. The rows of $\Psi$ are denoted by $\Psi_i, i \in [n]$.
Next, the $(r \times \alpha)$ message matrix $M$ is defined as 
$$ M = \begin{bmatrix} 
S_1 \\
S_2
\end{bmatrix}$$
where $S_1$ and $S_2$ are $(\alpha \times \alpha)$ symmetric matrices constructed such that ${k \choose 2}$ entries in the upper-triangular part of each matrix are filled up by ${k \choose 2}$ distinct message symbols and entries in the strictly lower-triangular are chosen to make the matrices symmetric. This is the MSR code we will use for our PIR scheme.

\section{System Model}

Here we formally describe the storage model and its retrieval scheme. Consider a storage system with $n$ non-colluding nodes that store a database $X$ consisting of $m$ records, each of length $\ell=k(k-1)$, denoted by $X^1, X^2, \dots, X^m \in \mathbb{F}_q^\ell$. Each record is encoded and distributed across $n$ nodes using the same product-matrix MSR code with parameters $$(n, k , r, \alpha, \beta, B) = (n,k,r,r-k+1,1,k(r-k+1)),$$ i.e. record $X^j$ is encoded by $$C^j = \Psi \cdot \mathcal{M}^j$$ where $\mathcal{M}^j$ is the corresponding message matrix of $X^j$. Write $$\mathcal{M} = \begin{bmatrix} 
\mathcal{M}^1 & \cdots & \mathcal{M}^m \\
\end{bmatrix},$$ and denote by $\mathcal{M}_i$ the row $i$ of $\mathcal{M}$. Hence, the whole system is $$C= \begin{bmatrix} 
C^1 & \cdots & C^m \\
\end{bmatrix} = \begin{bmatrix} 
\Psi \cdot \mathcal{M}^1 & \cdots & \Psi \cdot \mathcal{M}^m \\
\end{bmatrix} = \Psi \cdot \mathcal{M}$$ which means that each node stores $m\alpha = m(k-1)$ symbols in total, and we denote by $C_i$ the row $i$ of $C$ which consists of all symbols stored in node $i$, and $C_i^j$ the row $i$ of $C^j$ which consists of all symbols of $X^j$ stored in node $i$.

For the retrieval step, suppose the user wants to retrieve the record $X^{f}$. The user submits a $d \times m\alpha$ query matrix $Q^i$ over $GF(q)$ to node $i$, which we can think of $d$ rows of $Q^i$ as $d$ subqueries submitted to node $i$. In our scheme and in \cite{MDS}, for example, $d$ is set to be $k$. Then node $i$ responds with an answer $A_i^T = Q^i C_i^T$. The retrieval steps are as follows:

\begin{enumerate}[label=(\roman*)]
	\item (Initialisation) The user genarates an $d \times m\alpha$ matrix $U$ whose elements are chosen independently and uniformly at random over $GF(q)$. Let $U_j$ be row $j$ of $U$.
	\item (Query Generation) The query matrices are defined by a $d \times n\alpha$ matrix 
	$$ \mathbf{V} = \begin{bmatrix} 
	V^1 & \cdots & V^n \\
	\end{bmatrix}$$
	where $V^i$ is a $d \times \alpha$ binary matrix for $i \in [n]$. The request to each node $h$ is an $d \times m\alpha$ matrix $$Q^h = U + V^h E^f$$ where $$E^f = \left[\begin{array}{c|c|c}
	\textbf{0}_{\alpha \times (f-1)\alpha} & I_{\alpha \times \alpha} & \textbf{0}_{\alpha \times (m-f)\alpha}
	\end{array}\right].$$ In other words, $E^f$ is an $\alpha \times m\alpha$ matrix such that $C_h (E^f)^T = C_h^f$ which is a coded data piece of the desired record $X^f$ stored in node $h$. If the entry $(a,b)$ of $V^h$ is 1, then it implies that the entry $C^f_{hb}$ is privately retrieved by the $a^{th}$ subquery of $Q^h$.
	\item (Response Mappings) Each node $h$ returns $A_h^T = Q^h C_h^T$.
\end{enumerate}
Let $H(\cdot)$ be the entropy function. We say that a scheme is a \textit{perfect information-theoretic PIR scheme} if 
\begin{enumerate}
	\item
	(i)(privacy) $H(f|Q^i)=H(f)$ for every $i \in [n]$;
	
	\item
	(ii)(decodability) $H(X^f|A_1,\dots,A_n)=0$.
	
\end{enumerate}
According to our definition, (i) means that a node $i$ gets no information about which records are being retrieved by the user, and (ii) implies that having all responses $A_i$ from each node $i$ ensures that the user is able to recover the desired records $X^f$ with no errors.
\vspace{-10pt}\\

Two measurements are usually used to analyse the efficiency of the PIR scheme. First, \textit{Storage Overhead (SO)} is defined to be the ratio of the total storage used in the scheme to the total size of the entire database which, in our model, is $$SO = n(m\alpha) / mk\alpha = n/k.$$ In fact, storage overhead is the reciprocal of code rate. Second, the \textit{communication Price of Privacy (cPoP)} of a PIR scheme, which is defined in \cite{MDS}, is the ratio of the total amount of downloaded data to the total size of all desired records which, in our model, is $$cPoP = dn/k\alpha.$$

\noindent In this paper, we also consider the metric called the \textit{repair ratio (RR)} which is defined to be the ratio of the total amount of symbols downloaded for repairing a failed node to the size of the failed node which is equal to $$mr/m\alpha = r/(r-k+1)$$ in our model.

\section{Decodability Condition and Trade-off Analysis}
From the retrieval scheme, we can see that in fact, the response from node $i$ is
\begin{align*}
A_i &= C_i(Q^i)^T \\
&= C_i [U^T+(E^f)^T (V^i)^T] \\
&= C_i U^T + C_i (E^f)^T (V^i)^T \\
&= C_i U^T + (C_i^f)(V^i)^T.
\end{align*}
Then, the $j^{th}$ response in $A_i$ is $A_{ij} = C_i (U_j)^T + C_i^f (V^i_j)^T$ where $V^i_j$ is the row $j$ of $V^i$. Hence, record $X^f$ should be decoded by solving the system of linear equations $$A_{ij} = C_i (U_j)^T + C_i^f (V^i_j)^T,$$ for all $i \in [n], j \in [d]$ where the unknowns are $$(C_i (U_j)^T, i \in [n], j \in [d], C^f_{ab}, a \in [n], b \in [\alpha]).$$ Consider first the unknowns $C_i (U_j)^T, i \in [n], j \in [d]$, we can see that for each $j \in [d],$
\begin{align*}
C(U_j)^T &= \Psi \cdot \mathcal{M} \cdot (U_j)^T \\
&= \Psi \cdot \begin{bmatrix} 
I_1^j & \cdots & I_r^j \end{bmatrix}^T
\end{align*}
where $I_h^j = \mathcal{M}_h \cdot (U_j)^T, h \in [r].$
For the unknowns $C^f_{ab}, a \in [n], b \in [\alpha]$, we know that $$C^f_{ab} = \mbox{the entry }(a,b) \mbox{ of } \Psi \cdot \mathcal{M}^j,\mbox{ }\forall a \in [n], b \in [\alpha].$$ 
Hence, the retrieval scheme is decodable if the following system of linear equations
\[
\left\{
\begin{aligned} 
A_{ij} &= C_i (U_j)^T + C_i^f (V^i_j)^T, && \forall i \in [n], j \in [d] \\ 
C_s (U_t)^T &= \Psi_s \cdot \begin{bmatrix}
 I_1^t & \cdots & I_r^t \end{bmatrix}^T, && \forall s \in [n], t \in [d] \\ 
C^f_{ab} &= \mbox{the entry }(a,b) \mbox{ of } C^f, && \forall a \in [n], b \in [\alpha]  \end{aligned} 
\right.
\] has a unique solution, where the unknowns are $$( C_i (U_j)^T, i \in [n], j \in [d], C^f_{ab}, a \in [n], b \in [\alpha]).$$ This condition is called \textit{decodability condition}.
\\
\vspace{-5pt}

Next, we will give the trade-off analysis between storage overhead and cPoP. First, we count the number of unknowns in the system of linear equations in the decodability condition which is equal to $nd+n\alpha$. Next, we count the number of linearly independent equations in the system. Consider $$C_s (U_t)^T = \Psi_s \cdot \begin{bmatrix} I_1^t & \cdots & I_r^t \end{bmatrix}^T, \forall s \in [n], t \in [d],$$ so we have, for each $t \in [d]$, $$C(U_t)^T = \Psi \cdot \begin{bmatrix} 
I_1^t & \cdots & I_r^t \end{bmatrix}^T.$$ Since $\Psi$ is of rank $r$, it has a parity check matrix $P$ of rank $n-r$ such that $P \cdot \Psi = 0$. So we have $$P \cdot C(U_t)^T = P \cdot \Psi \cdot \begin{bmatrix} 
I_1^t & \cdots & I_r^t \end{bmatrix}^T = 0.$$ This gives us $n-r$ linearly independent equations for each $t \in [d]$. Then, for $$C^f_{ab} = \mbox{the entry }(a,b) \mbox{ of } C^f, \forall a \in [n], b \in [\alpha],$$ since any $k$ rows of $C^f$ would give us $\mathcal{M}^f$, the remaining $n-k$ rows must be able to be written in terms of linear combinations of those $k$ rows of $C^f$. This give us $(n-k)\alpha$ equations in $C^f_{ab}$. Hence, there are at most $nd+(n-r)d+(n-k)\alpha$ linearly independent equations in the system. If the retrieval scheme meets the decodability condition, then $$nd+n\alpha \leq nd+(n-r)d+(n-k)\alpha,$$ which implies that $$k\alpha \leq (n-r)d.$$ Therefore, $$1 \leq \frac{dn}{k\alpha}\bigg(1-\frac{r}{n}\bigg).$$ \\
In terms of storage overhead and cPoP we have $$1 \leq cPoP\bigg(1-\frac{r}{k(SO)}\bigg).$$ This shows that there is a trade-off between cPoP and storage overhead, and in terms of repair ratio and cPoP we have $$1 \leq cPoP - RR\bigg(\frac{d}{k}\bigg).$$ This shows that cPoP is bounded below by repair ratio.

\section{Our Construction}

In this construction, we use the product-matrix MSR code from \cite{matrix} with $n=3k-3$, over the finite field $\mathbb{F}_q$, so the parameters of the MSR code are $$(n,k,r,\alpha,\beta,B)=(3k-3,k,2k-2,k-1,1,k(k-1)).$$ We adapt the retrieval technique from the PIR schemes using MDS codes \cite{MDS} to our storage which uses the MSR codes where the desired record $X^f$ can be recovered from the coded data piece stored in any $k$ nodes, and we also make use of the repair property. We first start with an example to motivate our scheme.
\begin{example}
Suppose that we have 3 records over the finite field $\mathbb{F}_{13}$, each with size $6$, which can be written as $$X^i = \{x_{i1}, x_{i2}, x_{i3}, x_{i4}, x_{i5}, x_{i6}\}, \mbox{ for } i=1,2,3.$$
We use a $(6, 3, 4, 2, 1, 6)$ product-matrix MSR code over $\mathbb{F}_{13}$ to encode each record by choosing the encoding matrix $\Psi$ to be the Vandermonde matrix, and the message matrix $\mathcal{M}^i$ for the record $i , i \in \{1,2,3\}$ as described in \cite{matrix}:
\[ \Psi =
\begin{bmatrix}
1 & 1 & 1 & 1 & 1 & 1 \\
1 & 2 & 3 & 4 & 5 & 6 \\
1 & 4 & 9 & 3 & 12 & 10 \\
1 & 8 & 1 & 12 & 8 & 8
\end{bmatrix}^T, \quad  \mathcal{M}^i = \begin{bmatrix}
x_{i1} & x_{i2} \\
x_{i2} & x_{i3} \\
x_{i4} & x_{i5} \\
x_{i5} & x_{i6} 
\end{bmatrix}.
\]
Hence, each node stores
\begin{center}
	\scalebox{0.9}{
		\begin{tabular}{ |c|c|c| } 
			\hline
			node 1 & node 2 & node 3 \\
			\hline
			$x_{11}+x_{12}+x_{14}+x_{15}$ & $x_{11}+2x_{12}+4x_{14}+8x_{15}$ & $x_{11}+3x_{12}+9x_{14}+x_{15}$ \\ 
			$x_{12}+x_{13}+x_{15}+x_{16}$ & $x_{12}+2x_{13}+4x_{15}+8x_{16}$ & $x_{12}+3x_{13}+9x_{15}+x_{16}$ \\ 
			\hline
			$x_{21}+x_{22}+x_{24}+x_{25}$ & $x_{21}+2x_{22}+4x_{24}+8x_{25}$ & $x_{21}+3x_{22}+9x_{24}+x_{25}$ \\ 
			$x_{22}+x_{23}+x_{25}+x_{26}$ & $x_{22}+2x_{23}+4x_{25}+8x_{26}$ & $x_{22}+3x_{23}+9x_{25}+x_{26}$ \\ 
			\hline
			$x_{31}+x_{32}+x_{34}+x_{35}$ & $x_{31}+2x_{32}+4x_{34}+8x_{35}$ & $x_{31}+3x_{32}+9x_{34}+x_{35}$ \\ 
			$x_{32}+x_{33}+x_{35}+x_{36}$ & $x_{32}+2x_{33}+4x_{35}+8x_{36}$ & $x_{32}+3x_{33}+9x_{35}+x_{36}$ \\ 
			\hline
		\end{tabular}}
	\end{center}
	
\begin{center}
	\scalebox{0.9}{
		\begin{tabular}{ |c|c|c| } 
			\hline
			 node 4 & node 5 & node 6 \\
			\hline
			 $x_{11}+4x_{12}+3x_{14}+12x_{15}$ & $x_{11}+5x_{12}+12x_{14}+8x_{15}$ & $x_{11}+6x_{12}+10x_{14}+8x_{15}$\\ 
			 $x_{12}+4x_{13}+3x_{15}+12x_{16}$ & $x_{12}+5x_{13}+12x_{15}+8x_{16}$ & $x_{12}+6x_{13}+10x_{15}+8x_{16}$\\ 
			\hline
			 $x_{21}+4x_{22}+3x_{24}+12x_{25}$ & $x_{21}+5x_{22}+12x_{24}+8x_{25}$ & $x_{21}+6x_{22}+10x_{24}+8x_{25}$\\ 
			 $x_{22}+4x_{23}+3x_{25}+12x_{26}$ & $x_{22}+5x_{23}+12x_{25}+8x_{26}$ & $x_{22}+6x_{23}+10x_{25}+8x_{26}$\\ 
			\hline
		     $x_{31}+4x_{32}+3x_{34}+12x_{35}$ & $x_{31}+5x_{32}+12x_{34}+8x_{35}$ & $x_{31}+6x_{32}+10x_{34}+8x_{35}$\\ 
			 $x_{32}+4x_{33}+3x_{35}+12x_{36}$ & $x_{32}+5x_{33}+12x_{35}+8x_{36}$ & $x_{32}+6x_{33}+10x_{35}+8x_{36}$\\ 
			\hline
		\end{tabular}}
	\end{center} 
	Recall that $C^a_{ij}$ is the $j^{th}$ symbol of record $a$, stored in node $i$. Here the entire database can be recovered from the content of any 3 nodes, and if any one node failed, it can be repaired by downloading one symbol each from 4 of the remaining nodes.
\\
\vspace{-5pt}

	In the retrieval step, suppose the user wants record $X^1$. The query $Q^i$ is a $(3 \times 6)$ matrix which we can interpret as $3$ subqueries submitted to node $i$ for each $i \in [6]$. To form the query matrices, the user generates a $(3 \times 6)$ random matrix $U = [u_{ij}]$ whose entries are chosen uniformly at a random from $\mathbb{F}_{13}$. Choose  \[
	V^1 = \begin{bmatrix}
	1 & 0 \\
	0 & 1 \\
	0 & 0  
	\end{bmatrix}, \quad V^2 = \begin{bmatrix}
	0 & 0 \\
	1 & 0 \\
	0 & 1 
	\end{bmatrix}, \quad V^3 = \begin{bmatrix}
	0 & 1 \\
	0 & 0 \\
	1 & 0
	\end{bmatrix},
	\quad V^4 = V^5 = V^6 = \textbf{0}_{3 \times 2}.
	\]
	As $$E^1 = \begin{bmatrix}
	1 & 0 & 0 & 0 & 0 & 0 \\
	0 & 1 & 0 & 0 & 0 & 0 \\  
	\end{bmatrix},$$ we have
	\[
	V^1E^1 = \begin{bmatrix}
	1 & 0 & 0 & 0 & 0 & 0 \\
	0 & 1 & 0 & 0 & 0 & 0 \\
	0 & 0 & 0 & 0 & 0 & 0 
	\end{bmatrix}, V^2E^1 = \begin{bmatrix}
	0 & 0 & 0 & 0 & 0 & 0 \\
	1 & 0 & 0 & 0 & 0 & 0 \\
	0 & 1 & 0 & 0 & 0 & 0
	\end{bmatrix}, V^3E^1 = \begin{bmatrix}
	0 & 1 & 0 & 0 & 0 & 0 \\
	0 & 0 & 0 & 0 & 0 & 0 \\
	1 & 0 & 0 & 0 & 0 & 0
	\end{bmatrix},
	\]
	and $$V^4E^1 = V^5E^1 = V^6E^1 = \textbf{0}_{3 \times 6}.$$

	\noindent The query matrices are $Q^i = U+V^iE^1, i \in [6]$. Then each node computes and returns the length-$3$ vector $A_i^T = Q^i C^T_i$. Write $A_i = (A_{i1},A_{i2},A_{i3})$. Recall that 
	\begin{align*}
		\mathcal{M}_1 &= (x_{11}, x_{12}, x_{21}, x_{22}, x_{31}, x_{32}), \\
		\mathcal{M}_2 &= (x_{12}, x_{13}, x_{22}, x_{23}, x_{32}, x_{33}), \\
		\mathcal{M}_3 &= (x_{14}, x_{15}, x_{24}, x_{25}, x_{34}, x_{35}), \\
		\mathcal{M}_4 &= (x_{15}, x_{16}, x_{25}, x_{26}, x_{35}, x_{36}).
	\end{align*}
	Consider first subquery 1, we obtain
	\begin{align*}
		C^1_{11}+I^1_1+I^1_2+I^1_3+I^1_4 &= A_{11},  \tag{1}\\
		I^1_1+2I^1_2+4I^1_3+8I^1_4 &= A_{21},  \tag{2}\\
		C^1_{32}+I^1_1+3I^1_2+9I^1_3+I^1_4 &= A_{31},  \tag{3}\\
		I^1_1+4I^1_2+3I^1_3+12I^1_4 &= A_{41},  \tag{4}\\
		I^1_1+5I^1_2+12I^1_3+8I^1_4 &= A_{51},  \tag{5}\\
		I^1_1+6I^1_2+10I^1_3+8I^1_4 &= A_{61},  \tag{6}
	\end{align*}
	where $I^1_h = \mathcal{M}_h\cdot U_1^T, h=1,2,3,4$, and $U_1$ is the first row of $U$. 
	
	The user can solve for $I^1_1,I^1_2,I^1_3,I^1_4$ from $(2),(4),(5),(6)$ as they form the equation
	$$\begin{bmatrix}
	1 & 2 & 4 & 8 \\
	1 & 4 & 3 & 12 \\
	1 & 5 & 12 & 8 \\
	1 & 6 & 10 & 8
	\end{bmatrix} \cdot \begin{bmatrix}
	I^1_1 \\ I^1_2 \\ I^1_3 \\ I^1_4 
	\end{bmatrix} =  \begin{bmatrix}
	A_{21} \\ A_{41} \\ A_{51} \\ A_{61}
	\end{bmatrix}$$
	where the left matrix is the $(4 \times 4)$ submatrix of $\Psi$ which is invertible. Therefore, the user gets $C^1_{11}$ and $C^1_{32}$. Similarly, from subqueries 2, 3, the user obtains $C^1_{12}, C^1_{21}$ and $C^1_{22}, C^1_{31}$, respectively. Hence, the user has all the symbols of $X^1$ which are stored in the first $k=3$ nodes. From the recovery property of the regenerating code, the user can reconstruct $X^1$ as desired.
	\vspace{30pt}
	\begin{table}[h!]
		\centering
		\begin{tabular}{|c|c|c|c|c|c|} 
			\hline
			node 1 & node 2 & node 3 & node 4 & node 5 & node 6 \\
			\hline
			\cellcolor{yellow!25}1 & \cellcolor{cyan!25}2 & \cellcolor{magenta!25}3 &  &  & \\ 
			\hline
			\cellcolor{cyan!25}2 & \cellcolor{magenta!25}3 & \cellcolor{yellow!25}1 &  &  & \\ 
			\hline
		\end{tabular}
		\captionsetup{font=scriptsize, width=9.9cm}   
		\caption{Retrieval pattern for a $(6,3,4,2,1,6)$ MSR code. The $\alpha \times n$ entries correspond to $(C^f)^T$ and the entries labelled by the same number, say $d$, are privately retrieved by subquery $d$.}
		\label{table:1}
	\end{table}
\end{example}

Now we will give the general PIR scheme of our construction and prove the decodability and privacy. Recall that we use the MSR code with parameters $$(n,k,r,\alpha,\beta,B)=(3k-3,k,2k-2,k-1,1,k(k-1))$$ over $\mathbb{F}_q$ to store each record $X^1,\dots,X^m$, which means that $$C^i = \Psi \cdot \mathcal{M}^i$$ where $\mathcal{M}^i$ is the message matrix corresponding to $X^i$ as described in \cite{matrix}, so $$C = \begin{bmatrix}
\Psi\cdot\mathcal{M}^1 & \cdots & \Psi\cdot\mathcal{M}^m
\end{bmatrix}.$$
Suppose that the user wants record $X^f$. In the retrieval step, the user sends a $(k \times m\alpha)$ query matrix $Q^i$, which we can interpret as $k$ subqueries, to each node $i, i=1,\dots,n$. To form the query matrices, the user generates a $(k \times m\alpha)$ random matrix $U = [u_{ij}]$ whose entries are chosen uniformly at a random from $\mathbb{F}_q$. We choose $$V^1 = \left[\begin{array}{c}
 I_{(k-1) \times (k-1)} \\
 \textbf{0}_{1 \times (k-1)}
\end{array}\right]$$

\vspace{8pt} \noindent
and $V^j, j=2,\dots,k$ is obtained from matrix $V^{j-1}$ by a single downward cyclic shift of its row vectors. For $i=k+1,\dots,n$, we choose $V^i = \textbf{0}_{k \times (k-1)}.$ As $$E^f = \left[\begin{array}{c|c|c}
\textbf{0}_{\alpha \times (f-1)\alpha} & I_{\alpha \times \alpha} & \textbf{0}_{\alpha \times (m-f)\alpha}
\end{array}\right],$$ we have $$V^1E^f = \left[\begin{array}{cc|c|cc}
\multicolumn{2}{c|}{\smash{\raisebox{-0.5\normalbaselineskip}{$\textbf{0}_{k \times (f-1)\alpha}$}}} & I_{(k-1) \times (k-1)} &         &     \\
& & \textbf{0}_{1 \times (k-1)} &
\multicolumn{2}{c}{\smash{\raisebox{.5\normalbaselineskip}{$\textbf{0}_{k \times (m-f)\alpha}$}}}
\end{array}\right]$$

\vspace{8pt} \noindent
and $V^jE^f, j=2,\dots,k$ is obtained from matrix $V^{j-1}E^f$ by a single downward cyclic shift of its row vectors. For $i=k+1,\dots,n$, we have $V^i = \textbf{0}_{k \times m\alpha}.$ The query matrices are $Q^i = U+V^i, i \in [n].$ Then, each node computes and returns the length-$k$ $A^T_i = Q_i C^T_i$, and we write $A_i = (A_{i1},A_{i2},\dots,A_{ik})$.

\textit{Decodability:} The following proof will show the decodability of this scheme. We can see that for $i=1,\dots,n$, 
\begin{align*}
C_i &= \Psi_i \cdot \mathcal{M}\\
&= \Psi_i \cdot \begin{bmatrix}
\longdash \text{ } \mathcal{M}_1 \text{ } \longdash\\
\longdash \text{ } \mathcal{M}_2 \text{ }\longdash\\
\vdots \\
\longdash \text{ } \mathcal{M}_{2k-2} \text{ } \longdash\\
\end{bmatrix} \\
&= \sum_{j=1}^{2k-2} \Psi_{ij} \mathcal{M}_j
\end{align*}
Thus, $$C^T_i =  \sum_{j=1}^{2k-2} \Psi_{ij} \mathcal{M}_j^T.$$
\vspace{5pt} \\
\noindent Consider first subquery 1, we obtain
\begin{alignat*}{2}
C^f_{1,(f-1)\alpha+1}+\sum_{j=1}^{2k-2} \Psi_{1j} I^1_j  &= (U_1+e_{(f-1)\alpha+1})C^T_1 &&= A_{11},  \tag{1}\\
\sum_{j=1}^{2k-2} \Psi_{2j} I^1_j  &= U_1 C^T_2 &&= A_{21}, \tag{2}\\
C^f_{3,(f-1)\alpha+k-1}+\sum_{j=1}^{2k-2} \Psi_{3j} I^1_j  &= (U_1+e_{(f-1)\alpha+k-1})C^T_3 &&= A_{31},  \tag{3}\\
C^f_{4,(f-1)\alpha+k-2}+\sum_{j=1}^{2k-2} \Psi_{4j} I^1_j &= (U_1+e_{(f-1)\alpha+k-2})C^T_4 &&= A_{41},  \tag{4}\\
&\vdotswithin{=}   &&\vdotswithin{=} \\
C^f_{k,(f-1)\alpha+2}+\sum_{j=1}^{2k-2} \Psi_{k,j} I^1_j &= (U_1+e_{(f-1)\alpha+2})C^T_{k} &&= A_{k,1}, \tag{k}\\
\sum_{j=1}^{2k-2} \Psi_{k+1,j} I^1_j &= U_1 C^T_{k+1} &&= A_{k+1,1},  \tag{k+1} \\
&\vdotswithin{=}   &&\vdotswithin{=} \\
\sum_{j=1}^{2k-2} \Psi_{n,j}, I^1_j  &= U_1 C^T_{n} &&= A_{n,1},  \tag{n}
\end{alignat*}
where $I^1_h = \mathcal{M}_h \cdot U_1^T, h=1,2,\dots,2k-2$, $U_1$ is the first row of $U$, and $e_t$ is the length-$m\alpha$ binary unit vector with 1 at the $t^{th}$ position. 

The user can solve for $I^1_1,\dots,I^1_{2k-2}$ from $(2),(k+1),\dots,(n)$ as they form the equation
$$\begin{bmatrix}
\longdash \text{ } \Psi_2 \text{ } \longdash\\
\longdash \text{ } \Psi_{k+1} \text{ }\longdash\\
\vdots \\
\longdash \text{ } \Psi_n \text{ } \longdash\\
\end{bmatrix} \cdot \begin{bmatrix}
I^1_1 \\ I^1_2 \\ \vdots \\ I^1_{2k-2}
\end{bmatrix} =  \begin{bmatrix}
A_{21} \\ A_{k+1,1} \\ \vdots \\ A_{n,1}
\end{bmatrix}$$
where, since $n=3k-3$, the left matrix is $((2k-2) \times (2k-2))$ square submatrix of $\Psi$ which is invertible by the construction. Note that here we in fact make use of the repair property of the code, which requires $r \times r$ submatrices to be invertible. Therefore, the user gets $$C^f_{1,(f-1)\alpha+1},C^f_{3,(f-1)\alpha+k-1}, C^f_{4,(f-1)\alpha+k-2}, \dots, C^f_{k,(f-1)\alpha+2},$$ i.e., all the symbols with label 1 in Table 2. Similarly, from subqueries $i=2,\dots,k$, the user obtains all the symbols with label $i$ in Table 2. Hence, the user has all the symbols of $X^f$ which are stored in the first $k$ nodes. From the recovery property of the regenerating code, the user can finally reconstruct $X^f$ as desired.
\\
\vspace{-5pt}

\textit{Privacy:} As we construct the query matrices $Q^i$ via the random matrix $U$, $Q^i$ is independent from $f$ which implies that this scheme achieves perfect privacy.

\vspace{10pt}
\begin{table}[h!]
	\centering
	\begin{tabular}{ |c|c|c|c|c|c|c|c|c| } 
		\hline
		node 1 & node 2 & node 3 & $\cdots$ & node $k-1$ & node $k$ & node $k+1$ & $\cdots$ & node $n$ \\
		\hline
		\cellcolor{yellow!25}1 & \cellcolor{cyan!25}2 & \cellcolor{magenta!25}3 & $\cdots$ & \cellcolor{blue!25}$k-1$ & \cellcolor{black!25}$k$ &  & & \\ 
		\hline
		\cellcolor{cyan!25}2 & \cellcolor{magenta!25}3 & \cellcolor{green!25}4 & $\cdots$ & \cellcolor{black!25}$k$ & \cellcolor{yellow!25}1 &  & & \\ 
		\hline
		$\vdots$ & $\vdots$ & $\vdots$ & $\vdots$ & $\vdots$ & $\vdots$ & & &  \\
		\hline
		\cellcolor{red!25}$k-2$ & \cellcolor{blue!25}$k-1$ & \cellcolor{black!25}$k$ & $\cdots$ & \cellcolor{blue!50}$k-4$ & \cellcolor{magenta!50}$k-3$ & & & \\
		\hline
		\cellcolor{blue!25}$k-1$ & \cellcolor{black!25}$k$ & \cellcolor{yellow!25}1 & $\cdots$ & \cellcolor{magenta!50}$k-3$ & \cellcolor{red!25}$k-2$ &  & & \\
		\hline
	\end{tabular}
	\caption{Retrieval pattern}
	\label{table:2}
\end{table}

\section{Discussion and Future Work}

In our PIR scheme, storage overhead is $$\frac{n}{k}=\frac{3k-3}{k},$$
and cPoP equals to $$\frac{dn}{k\alpha} = \frac{k(3k-3)}{k(k-1)} = 3,$$ Hence, 
\begin{align*}
cPoP\bigg(1-\frac{r}{k(SO)}\bigg) &= 3\bigg(1-\frac{2k-2}{3k-3}\bigg) \\
&= 3\bigg(1-\frac{2}{3}\bigg) \\
&= 1.
\end{align*}
This means that our scheme achieves the information theoritic limit as it fits the optimal curve in the trade-off derived in Section 4.

Also, as we use the MSR codes in our construction which beneficially reduces the repair cost when a node failure occurs in the system, repair ratio in our scheme is $$\frac{r}{\alpha} = \frac{(2k-2)}{k-1} = 2,$$ which is smaller than PIR schemes that use $(n,k)$-MDS codes (for example, in \cite{MDS}) where the repair ratio equals to $k$ if $k>2$.

We remark that a similar construction still works for product-matrix minimum bandwidth regenerating (MBR) codes with some adjustment which we will publish later. Also, the construction of schemes using this retrieval technique for the Multi-message PIR, which is a variation of PIR
that allows a user to download multiple messages from the database without revealing the identity of the desired messages, is achievable and we will present this at the 2018 IEEE International Symposium on Information Theory (ISIT) at the Recent Results Session.

\end{document}